\documentclass[aps,prl,preprint,groupedaddress]{revtex4-1}

\usepackage{graphicx}

\begin{document}


\title{Lasing in localized modes of a slow light photonic crystal waveguide}


\author{Jin-Kyu Yang$^{1,2}$, Heeso Noh$^1$, Michael J. Rooks$^1$, Glenn S. Solomon$^3$, Frank Vollmer$^{4,5}$, Hui Cao$^1$}


\affiliation{$^1$ Department of Applied Physics, Yale University, New Haven, CT 06520-8482, USA}

\affiliation{$^2$ Department of Optical Engineering, Kongju National University, Kongju 314-701, Republic of Korea}

\affiliation{$^3$ Joint Quantum Institute, NIST and University of Maryland, Gaithersburg, Maryland 20899, USA} 

\affiliation{$^4$ Rowland Institute at Harvard, Harvard University, Cambridge, Massachusetts 02142, USA}

\affiliation{$^5$ Max Planck Institute for the Science of Light, Erlangen, Germany}


\date{\today}

\begin{abstract}
We demonstrate lasing in GaAs photonic crystal waveguides with InAs quantum dots as gain medium. Structural disorder is present due to fabrication imperfection and causes multiple scattering of light and localization of light. Lasing modes with varying spatial extend are observed at random locations along the guide. Lasing frequencies are determined by the local structure and occur within a narrow frequency band which coincides with the slow light regime of the waveguide mode. The three-dimensional numerical simulation reveals that the main loss channel for lasing modes located away from the waveguide end is out-of-plane scattering by structural disorder.  
\end{abstract}


\maketitle


Light-matter interactions can be greatly enhanced by reducing the group velocity of light propagating in the materials. The slow light has applications to compact optical switches, modulators and low-threshold lasers \cite{NPT08Baba}. Photonic-crystals are attractive for generating slow light, as they are compatible with on-chip integration and room-temperature operation. The most common implementation is a photonic crystal waveguide (PCW) formed by a line defect. The defect band dispersion can be engineered to produce a vanishing group velocity. However, structural imperfection introduced unintentionally during the fabrication process causes a significant loss for light propagating in the slow light PCW \cite{PRB05Kuramochi,JLT08Parini,OE05Skorobogatiy}. In the absence of material absorption, the loss comes from backscattering and out-of-plane scattering of light by the structural disorder \cite{PRL05Hughes,APL04Povinelli,OL04Gerace,OE07Ofaolain,PRL08Engelen,PRB08Wang,PRL09Mazoyer}. The slower the light travels, the stronger the effect of disorder. Multiple scattering in the PCW may induce light localization, which has been studied both experimentally and theoretically \cite{PRL07Topo,APL07Topo,PRL09Patt,OE09Topo,PRB10Garcia}.  

Despite of its detrimental effect on light propagation through a slow light PCW, structural disorder can dramatically enhance spontaneous emission of light \cite{PRL08Hansen,APL10Thyr,Sci10Sapi}. It results in the formation of strongly confined modes with high quality ($Q$) factor and small modal volume, which the emitters are coupled to.  Although enhanced photoluminescence has been observed, lasing in a disordered PCW has not been realized \cite{APL08Yang}. In this paper, we demonstrate lasing in the localized modes of a slow light PCW with remnant structural disorder that is fabricated on a GaAs membrane. InAs quantum dots (QDs) embedded in the GaAs layer provide gain under optical pumping. The PCW is formed by removing a single row in a triangular lattice of air holes etched to the GaAs membrane. The fabrication imperfection causes random variation of air hole size and shape. Lasing modes are spatially localized, and their frequencies shift across the PCW due to a gradual variation of air hole size from the proximity effect. Three dimensional (3D) numerical simulation of real samples captures the position dependent lasing characteristic. 

\begin{figure}
\includegraphics[width=8cm]{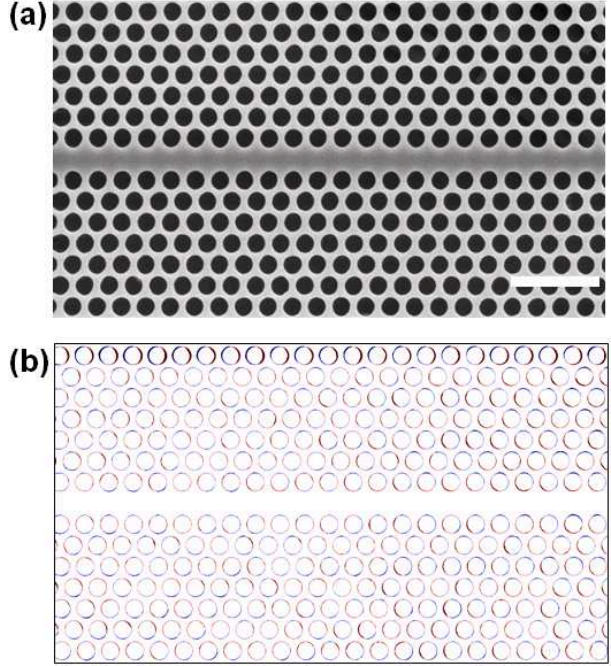}
\caption{(a) Top-view SEM image of part of a fabricated 2D PCW in a free-standing GaAs membrane. The white scale bar is 1 $\mu$m. The lattice constant $a$ = 280 nm and the radius of air hole $r$ = 0.37 $a$. (b) Difference between the digitized SEM image and the perfect photonic crystal with one missing row of air holes reveals the structural disorder that is introduced unintentionally during the fabrication process.}
\label{WGsem}
\end{figure}

We fabricated PCWs without any intentional disorder on a free-standing GaAs membrane. 190 nm GaAs and 1000 nm Al$_{0.75}$Ga$_{0.25}$As were grown on a GaAs substrate by molecular beam epitaxy. Three monolayers of InAs QDs equally spaced by 25 nm GaAs barriers were embedded in the GaAs layer. The PCW pattern was written on a 300-nm-thick ZEP layer by commercial electron-beam lithography. Then the pattern was transfered into the GaAs layer by chlorine-based inductive coupled plasma reactive ion etching (ICP-RIE) with the ZEP layer as the mask. The ZEP layer was subsequently removed in an oxygen plasma cleaning process. Finally the Al$_{0.75}$Ga$_{0.25}$As layer was selectively etched in a 10$\%$ dilute HF solution. The lattice constant $a$ was varied from 240 nm to 300 nm with 20 nm step, and the radius of air hole $r$ is fixed at 0.37 $a$. Figure \ref{WGsem}(a) is a top-view scanning electron microscope (SEM) image of a fabricated PCW. 
Figure \ref{WGsem}(b) shows its deviation from the designed pattern. Among the disorder in size, shape and position of air holes, the shape disorder dominates, and it is generated by nano-patterning and etching process. 

In the lasing experiment, the sample was mounted in a continuous-flow helium cryostat and cooled to 10 K,  because the emission efficiency of InAs QDs strongly depends on temperature. A mode-locked Ti:Sapphire (center wavelength = 790 nm, repetition rate = 76 MHz, pulse width = 200 fs) was used to excite the InAs QDs. The pump beam was focused normally onto the PCW through a  long-working-distance objective lens (numerical aperture = 0.40). The diameter of the pump spot on the sample was about 2 $\mu$m. The emission scattered out of plane was collected by the same objective lens and routed to a half-meter spectrometer with a liquid-nitrogen-cooled CCD array detector. The spectral resolution is 0.03 nm. The pump spot was moved along the PCW to probe lasing at different locations. 

\begin{figure}
\includegraphics[width=12cm]{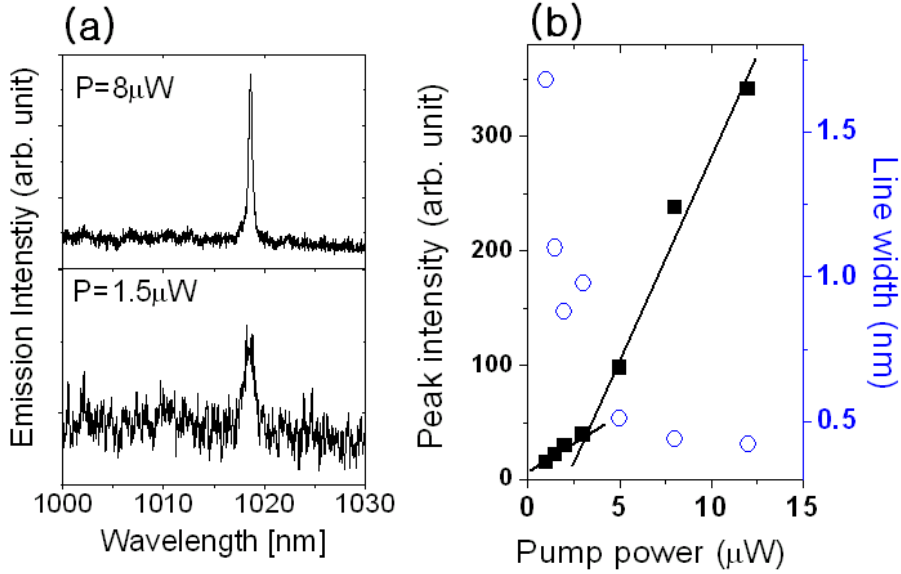}
\caption{(a) Measured emission spectra at the incident pump power $P$ = 1.5 $\mu$W (bottom) and 8.0 $\mu$W (top). (b) Intensity (black solid squares) and spectral width (blue open circles) of the emission peak in (a)  vs. incident pump power $P$. The lasing threshold $P \sim 3$ $\mu$W is highlighted by the two straight lines. }
\label{WGlasing}
\end{figure}

Figure \ref{WGlasing} displays the typical lasing characteristic at a fixed pump position. The emission spectra in Fig. \ref{WGlasing}(a) features a single peak. Its intensity exhibits a threshold behavior in Fig. \ref{WGlasing}(b). When the incident pump power $P$ exceeds 3 $\mu$W, the peak intensity grows much faster. The spectral width $\Delta \lambda$ of the peak also decreases dramatically with increasing $P$. Well above the threshold, the hot carrier effect produced by the short pump pulse with high peak power prevents a further reduction of $\Delta \lambda$ \cite{APL10Yang}. 

\begin{figure}
\includegraphics[width=10cm]{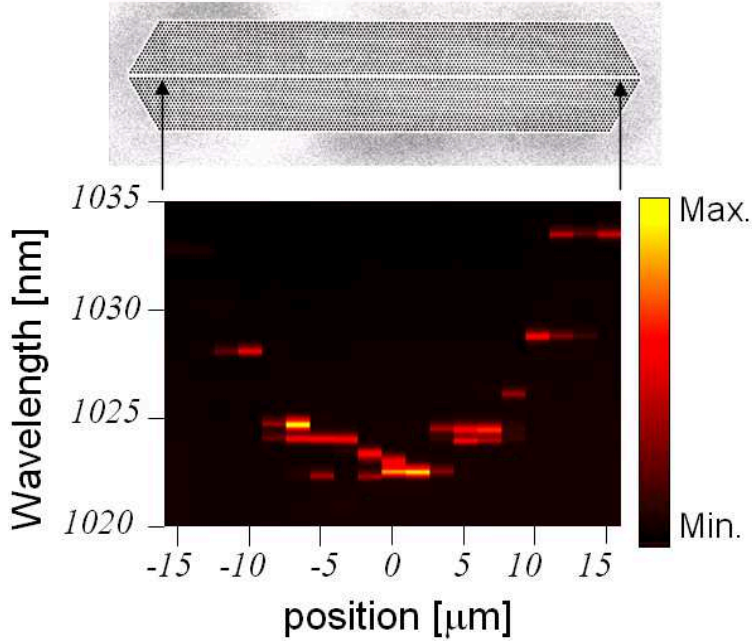}
\caption{Contour plot of position-dependent lasing spectra with local pumping. The horizontal axis is pumping position $x$ and the  vertical axis is wavelength $\lambda$. The left bar indicates the color code of emission intensity. The upper inset is a SEM image of the whole PCW of length 38 $\mu$m.}
\label{pos}
\end{figure}

Next we scanned the pump spot across the PCW to probe lasing modes at different locations. The pump spot of diameter $\sim$ 2 $\mu$m was moved in 2-$\mu$m step along the row of missing air holes. The pump power was fixed at a value above the lasing threshold. Figure \ref{pos} is a map of position-dependent lasing spectra with local pumping. The horizontal axis is pumping position $x$, and the vertical axis is wavelength $\lambda$. The emission intensity is color coded (see the right bar). On the top is a SEM image of the entire PCW of length 38 $\mu$m. As the pump spot was moved from the left end to the right end of the PCW, the lasing peaks first blue shift and then red-shift. Such shift was caused by the proximity effect of electron beam lithography. Since we used the positive electron beam resist, the center part of the pattern was over-exposed by backscattering of injected electrons from the GaAs substrate. Therefore, the air hole size in the center was slightly larger than that at the edge. Although the proximity-effect correction method was adopted in the writing process, such effect was not completely removed. The lasing spectra in Fig. \ref{pos} also illustrate that the shift of lasing frequency was discontinuous. Namely, as the pump spot was moved, the existing lasing peaks disappeared and new lasing peaks with distinct frequencies appeared. This means the lasing modes are spatially localized and their frequencies are determined by the local disorder. At some pump locations, e.g. $x$ = -7 $\mu$m, 5 $\mu$m, multi-mode lasing was observed.  The spatial spread of individual lasing modes could be quite different. For example, the lasing mode at $\lambda$ = 1024 nm extended from $x$ = -8 $\mu$m to $x$ = -3 $\mu$m. Its confinement along the guide was estimated to be approximately 6 $\mu$m. However, the lasing mode at $\lambda$ = 1026 nm and $x$ = 10 $\mu$m disappeared when the pump spots was moved 2 $\mu$m to the left or the right. Thus the mode size along the waveguide was 2 $\mu$m or less. We also observed a coupled mode at $\lambda$ = 1033 nm with two intensity maxima at $x$ = 12 $\mu$m and 16 $\mu$m. 
 
\begin{figure}
\includegraphics[width=10cm]{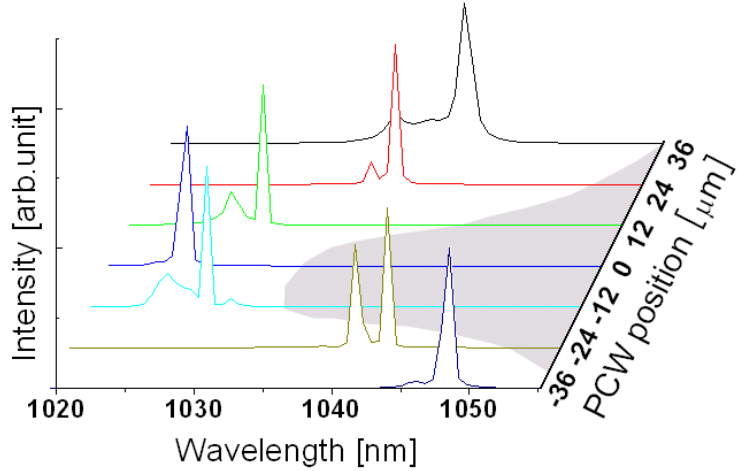}
\caption{Numerical simulation of local excitation spectra showing the frequencies of high-$Q$ modes vary with the position along the PCW. The gray region indicates the lowest-order waveguide band.}
\label{figcal}
\end{figure}

To understand the experimental data, we performed numerical simulation of the real structures extracted from the digitized SEM images. First we calculated the band structure of PCW using the plane-wave-expansion method \cite{MPB}. The measured lasing frequencies are close to the edge of the lowest-order waveguide band within the gap of the triangular lattice.  In this frequency regime the group velocity is reduced. Hence, the lasing modes are in the slow light regime. Next we used the 3D finite-difference time-domain (FDTD) method to calculate the high-quality ($Q$) modes in the passive PCW \cite{Sci04Park, APL10Yang}. Such modes have low lasing threshold, and correspond to lasing modes when optical gain is introduced. The spatial resolution of our numerical calculation was about $a/20$. The roughness of the sidewalls of air holes was ignored.  A broad-band seed pulse was launched from a specific position in the PCW. The optical spectrum taken long after the excitation pulse was gone contained only the high-$Q$ modes in the proximity of the source. 
Figure \ref{figcal} shows a series of such spectra when the excitation pulse was launched from different positions of the PCW.  The high-$Q$ modes shift to shorter wavelength as the source was moved from either end of the waveguide to the center. This trend matched well to the experimental observation. Since the average size of air holes increases towards the center of PCW, the waveguide band edge moves to higher frequency \cite{PRL07Topo}, causing a blue shift of the high-$Q$ modes. The wavelengths of high-$Q$ modes near the PCW center agree well to those of lasing modes observed experimentally. The electromagnetic field distributions of the high-$Q$ modes confirm that they are spatially localized not only in the direction perpendicular to the waveguide but also parallel to it. The quality factor of the modes near the PCW center is significantly higher than those near either end because of stronger in-plane confinement. We computed the energy loss both in-plane and out-of-plane. For the high-$Q$ modes at the PCW center, the loss ratio (the ratio of the out-of-plane loss and the in-plane loss is about 2.8. In contrast, the modes near the PCW end has a very small loss ratio about 0.14. It means that the mode near the PCW end has stronger in-plane loss due to light leakage from the end of the waveguide.  

In conclusion, we realized lasing in slow light photonic crystal waveguides with optical pumping. The structural disorder, created by fabrication imperfection, causes multiple scattering of light that leads to spatial localization of lasing modes. The localization length varies from mode to mode. The lasing frequencies are within the slow light regime and determined by the local structure. The three-dimensional numerical simulation reveals that the main loss channel for lasing modes located away from the waveguide end is out-of-plane scattering by structural disorder.  

We thank Stephen Hughes and Patrick Sebbah for useful discussions. This work is funded by NSF Grant No. DMR-0808937. J.-K. Yang acknowledges the support of KOSEF through the Grant for the Photonics Integration Technology Research Center (R11-2003-022) at OPERA.

\end{document}